# Position Dependence of Charge Collection in Prototype Sensors for the CMS Pixel Detector

Tilman Rohe, *Member, IEEE,* Daniela Bortoletto, Vincenzo Chiochia, Lucien M. Cremaldi, *Member, IEEE,*
Susanna Cucciarelli, Andrei Dorokhov, Marcin Konecki, Kirill Prokofiev, Christian Regenfus,
David A. Sanders, *Member, IEEE,* Seunghee Son, Thomas Speer, Morris Swartz

*Abstract*— This paper reports on the sensor R&D activity for the CMS pixel detector. Devices featuring several design and technology options have been irradiated up to a proton fluence[1] of $1 \times 10^{15}$ $n_{eq}/cm^2$ at the CERN PS. Afterward they were bump bonded to unirradiated readout chips and tested using high energy pions in the H2 beam line of the CERN SPS. The readout chip allows a non zero suppressed full analogue readout and therefore a good characterization of the sensors in terms of noise and charge collection properties. The position dependence of signal is presented and the differences between the two sensor options are discussed.

## I. INTRODUCTION

THE CMS experiment, currently under construction at the Large Hadron Collider (LHC) at CERN (Geneva, Switzerland), will contain a hybrid pixel detector for tracking and vertexing. It will consist in its final configuration of three barrel layers and two end disks at each side. The barrel layers will be 53 cm long and will have radii of 4.3 cm and 7.2 cm, while the third layer at 11.0 cm will be added later to provide a 3 layer system. The end disks are located at a mean distance to the interaction point of 34.5 cm and 46.5 cm. The whole system will provide three high resolution space points up to a pseudorapidity[2] of $|\eta| < 2.2$.

In order to achieve the best vertex position measurement the spatial resolution of the sensor should be as good in the $z$-direction (parallel to the beam line) as in $(r, \phi)$ and therefore almost a squared pixel shape with a pitch of $100 \times 150$ µm$^2$ was adopted. To improve the spatial resolution analog interpolation between neighboring channels will be performed. The strong Lorentz deflection in the $(r, \phi)$-direction caused by CMS' 4 T magnetic field is used to distribute the signal over two and more pixels. Hence the detectors are not tilted in the barrel layers. The resolution along the $z$-axis is determined by the pixel pitch in the region with low pseudorapidity and by charge sharing if the tracks hit the sensors at an inclined angle, the cluster size can exceed 6 or 7. The best resolution will be reached at the point where the charge is distributed over two pixels. In the disks, where the charge carrier drift is hardly affected by the magnetic field, the modules are tilted by about $20^o$ resulting in a turbine like geometry. This paper reports on the development of the sensor part of the system. A general overview on the CMS pixel project is given in [1].

Because of the harsh radiation environment at the LHC, the technical realization of the pixel detector is extremely challenging. The irradiation induced effects in silicon can be divided into surface and bulk damage. The oxide charge increases until its saturation value of a few $10^{12}$ cm$^{-2}$ reached after some kGy [2]–[4]. The concentration of interface traps also increases. Both effects influence the electric fields close to the surface and have to be considered when designing a sensor for radiative environments.

The leakage current increases in proportion to the hadron fluence [5], [6]. At a hadron fluence of about $5 \times 10^{13}$ $n_{eq}/cm^2$ the space charge in the depletion zone converts from positive (*n*-type) to negative ("*p*-type"). At higher fluence it increases proportionally to the fluence [5], [7]. The change of the effective doping concentration shows a complex annealing behavior with exponential dependence on the sensor's temperature [8], [9]. The change of material parameters with irradiation can be influenced by adding impurities to the silicon starting material [10]. A high oxygen content in the starting material acts beneficial in two ways. The introduction rate of positively charged defects is reduced and reverse annealing is slowed down. The CMS pixel detector will be operated at a temperature of about $-10^o$ C [1] to suppress the increasing leakage current and also kept cool outside data taking periods to suppress reverse annealing.

Trapping of the drifting charge in traps with emission time longer than the shaping time of the readout electronics of about 25 ns will reduce the signal. This effect is described by the effective trapping time constant which is inversely proportional to the hadron fluence [5], [11]. Trapping will eventually limit the use of silicon detectors for fluences exceeding $1 \times 10^{15}$ $n_{eq}/cm^2$.

The innermost barrel layer will be exposed to a fluence of about $3 \times 10^{14}$ $n_{eq}/cm^2$ per year at the full LHC-luminosity, the 2$^{nd}$ and 3$^{rd}$ layer to about $1.2 \times 10^{14}$ $n_{eq}/cm^2$ and $0.6 \times 10^{14}\,\mathrm{n_{eq}/cm^2}$, respectively. All components of the pixel detector are specified to remain operational up to a particle fluence of at least $6 \times 10^{14}\,\mathrm{n_{eq}/cm^2}$. This implies that the 2$^{nd}$ layer will have to be replaced once after about 7 years (3 with reduced and 4 with full luminosity) while the innermost layer will probably be replaced every 1 to 2 years of full luminosity equivalent.

The pixel sensors have to deliver a sufficiently high signal until the specified fluence. The final readout chips feature built-in data sparsification with a threshold set to 2000-3000 electrons in order to suppress noise hits. With a sensor thickness of $285\,\mathrm{\mu m}$, a minimum ionizing particle creates about 22 000 electron-hole pairs (most probable value). However, with increasing irradiation this charge cannot be fully collected due to trapping and incomplete depletion. As both effects can be reduced by increasing the sensor bias, the choice of the sensor concept must allow the application of elevated bias voltages without causing electrical breakdown. For the CMS pixel detector a maximum value of 500-600 V is foreseen.

In addition to the radiation-induced bulk effects the pixel design (i.e. the implant geometry) influences the charge collection properties of the sensor. The pixel-design has to be optimized to minimize potential regions of reduced signal collection. The aim of this study is to compare two options for the sensors of the CMS pixel detector with respect to their signal collection properties. Therefore the position dependence of the collected signal within every pixel is evaluated.

## II. Sensor Concepts under Study

After the irradiation-induced space charge sign inversion of the substrate and the subsequent increase of the full depletion voltage, sensors might have to be operated partially depleted. Therefore an "$n$-in-$n$" concept has been chosen. In addition double-sided processing of these devices allows the implementation of guard rings only on the $p$-side of the sensor, keeping all sensor edges at ground potential. The design of the guard rings has been optimized in the past [12]. The breakdown voltage exceeds by safely the required value of 600 V.

In order to detect the signal on the ohmic $n$-side of the sensor inter-pixel isolation has to be provided. Here $p$-stops are considered as well as the $p$-spray technique. The pixel-layouts of the two design options investigated for this study are shown in Fig. 1. In order to test the segmented devices on wafer with current-voltage (IV) measurements and to keep accidentally unconnected pixel cells close to ground potential, high resistive electrical connections between the pixels have been implemented. In the case of $p$-stops this was realized by openings in the $p$-stop implants. The fixed positive oxide charge builds up a electron accumulation that forms a "resistive network" to which all pixels are connected via the openings (see Fig. 2). The properties of such "resistors" have been studied in detail in [12], [13]. According to [13], [14] the most promising geometries feature small distances between the $n^+$-implants and quite large $p$-stop openings. Both are realized in the design under study shown in Fig.1a.

In addition we investigated prototypes featuring the moderated $p$-spray isolation technique. Here the isolating $p$-implant is performed without a photo-lithographic mask and therefore no structuring is possible. However punch-through biasing can be implemented. Its behavior is much less dependent on external conditions like backside bias and radiation effects than the resistors formed by the electron accumulation. The layout (see Fig. 1b) is characterized by small gaps of $20\,\mathrm{\mu m}$ between the $n^+$-implants and by a minimized biasing structure using small "bias dots" [15].

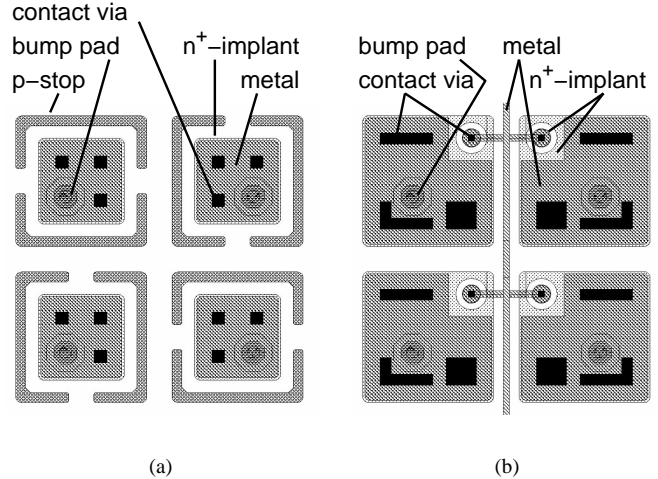

Fig. 1. Mask layout of the pixel sensors under study. Open $p$-stop rings (a) and $p$-spray with bias grid (b).

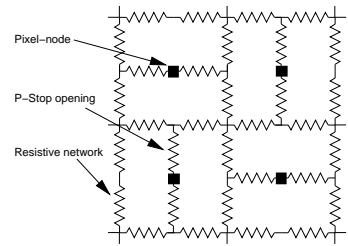

Fig. 2. Sketch of the "resistive network" formed by the electron accumulation layer and the $p$-stop openings. Each pixel node is connected to it by the openings in in the $p$-stop rings.

The pixel size of the sensors investigated in this study was $125 \times 125\,\mathrm{\mu m^2}$ in order to match the readout chip. Although these dimensions differ slightly from the cell size used in CMS we are confident that the basic charge collection properties presented in this paper are not affected by the change of the cell size to $100 \times 150\,\mathrm{\mu m^2}$. Other properties, as for example the spatial resolution, have to be measured with the final configuration.

Following the recommendation of the ROSE collaboration [10], oxygen enriched silicon was used in this prototypes to improve the post irradiation behavior.

## III. Test Procedure

The results presented in sections IV-B – D were obtained with sensors containing $22 \times 32$ pixels. In a pixelated device the parameters important for the performance of a single channel, like pixel capacitance and leakage current, are independent of the array dimensions. Therefore the use of miniature sensors does not restrict the validity of the results.

After the deposition of the under bump metalization and the indium bumps the sensors were cut out of the wafers. Some of them were irradiated at the CERN PS with $24\,\text{GeV}$ protons (hardness factor 0.62 [10]). The irradiation was performed without cooling and bias. The fluences applied were 3, 8, and $11 \times 10^{14}\,\text{n}_{eq}/\text{cm}^2$. The sensors irradiated to $8 \times 10^{14}\,\text{n}_{eq}/\text{cm}^2$ were used to compare the performance of the two different types of sensor. In order to avoid reverse annealing the sensors were stored at $-20^o\,\text{C}$ after irradiation and warmed up only for transport and bump bonding. For the irradiated sensors a special bump bonding procedure without heat application was used. The total time without cooling was kept a short as possible and therefore all devices are annealed close to the minimum of the full depletion voltage. Prior to bump bonding all sensors were characterized with IV-measurements.

Several miniature sensors of the two designs were bump bonded to readout chips of the type PSI30/AC30[3] described in detail in [16]. This chip was chosen instead of the final CMS-pixel readout chip, fabricated in a $0.25\,\mu\text{m}$ standard CMOS-process, because it allows one to force a sequential readout of all 704 pixel cells without zero suppression. In order to readout the full chip, all comparators were switched off by masking the pixels and setting the thresholds to very high values. The sampling time at the shaper was defined by an external hold signal. In the test beam setup a pin-diode was used to provide the external hold signal and to trigger the readout.

The peaking times of the preamplifier and the shaper were adjusted to about $40\,\text{ns}$ by tuning the feedback resistors of the charge sensitive amplifiers. This setting prevents saturation of the preamplifier and shaper up to signals corresponding to about 1.5 minimal ionizing particles (m.i.p.) but leads to a higher noise.

The bump bonded samples were tested at the CERN-SPS H2 beam line using $150 - 225\,\text{GeV}$ pions. The pixel device under test was situated in-between a four layer silicon strip telescope [17] with an intrinsic spatial resolution of about $1\,\mu\text{m}$. The whole set-up was placed in a $3\,\text{T}$ magnet with the $\vec{B}$ field parallel to the beam. The pixel detector was set either normal to the beam, or with an angle of $15^o$ between the beam and the sensor surface. As this paper discusses the differences between the designs, it only reports on data taken at normal incidence without magnetic field. The measurements performed in magnetic field and with a tilted sensor were used to determine bulk properties like the Lorentz angle [18]. The irradiated sensors were operated at $-20^o\,\text{C}$ by the means of water cooled Peltier elements.

## IV. RESULTS

### A. Inter-Pixel Resistance

While $p$-spray isolated devices naturally feature high inter-pixel isolation the inter-pixel resistance of the open $p$-stop devices depends very much on the $p$-stop geometry and the width of the openings. In addition the quality of the oxide and the crystal orientation have a strong impact on the resistance value. In order to define a minimum operation voltage and to understand the charge charing of the open $p$-stop sensors, the inter-pixel resistance was measured in the unirradiated state.

[3]PSI30 DMILL pixel readout chip was designed in 1997 at Paul Scherrer Institut, Villigen, Switzerland, and translated in 1998 to the Honeywell RICMOS IV process at 1. Physikalisches Institut of the RWTH Aachen, Germany.

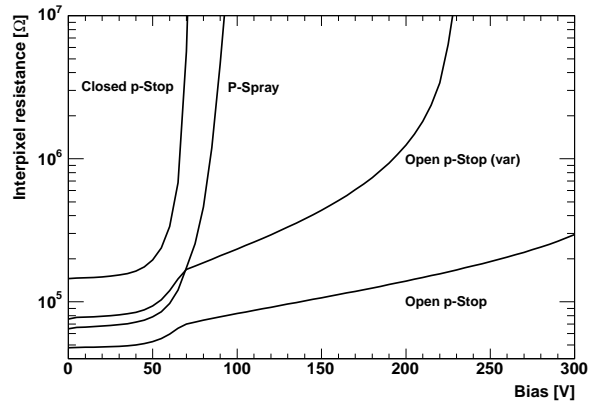

Fig. 3. Resistance between one pixel and all its neighbors as a function of the back side voltage. For comparison to the designs in Fig. 1 (open $p$-stop and $p$-spray) two other design options (open $p$-stop (var) and closed $p$-stop) are also shown.

For irradiated sensors this measurement was not performed as the resistance of the electron accumulation layer was found to be much higher (in the order of $G\Omega$) and was almost independent on the geometry [12], [13].

Fig. 3 shows the inter-pixel resistance of different pixel designs as a function of the sensor bias. The measurement was performed using a special test structure containing a grounded array of $5 \times 5$ pixels. The potential of the center pixel of the array was set to $+1\,\text{V}$. The current flowing into this pixel was measured as a function of the back side voltage. In order to illustrate the effect of the $p$-stop openings in the open $p$-stop design (Fig. 1a), an identical design with closed $p$-stops, a different open $p$-stop geometry ("open $p$-stop (var)") and the $p$-spray design (Fig. 1b) are also shown.

As the depletion starts from the back side ("$n$-in-$n$"), part of the current flows through the bulk before full depletion is reached and the corresponding inter-pixel resistance is low. With progressing depletion this channel is pinched off and the resistance in the fully isolated devices increases rapidly by several orders of magnitude. In the devices featuring $p$-stop openings a residual current flows over the electron accumulation layer. However, with the backside bias being increased further, this electron channel also starts to be pinched off. This is visible in Fig. 3 for the curve labeled "open $p$-stop (var)" at bias above $200\,\text{V}$.

The design shown in Fig. 1a (open $p$-stop) shows no pinch off up to $300\,\text{V}$. Its inter-pixel resistance at backside voltage of $150\,\text{V}$ is only about $100\,\text{k}\Omega$. This results in wide signal spreading along the resistive channels. The test beam data with the unirradiated device was therefore taken at $300\,\text{V}$ bias voltage where the inter-pixel resistance reaches a value of $300\,\text{k}\Omega$. At this bias voltage we observe a reduced but still notable spread of the collected charge over several pixels.

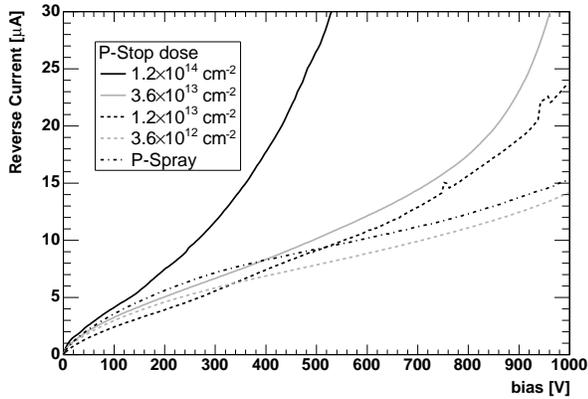

Fig. 4. IV curves of open *p*-stop sensors (Fig. 1a) with different *p*-stop implantation doses irradiated to $\Phi = 8 \times 10^{14}\,\mathrm{n_{eq}/cm^2}$, measured at $-20^o$ C. For comparison a *p*-spray sensor (Fig. 1b) is also plotted.

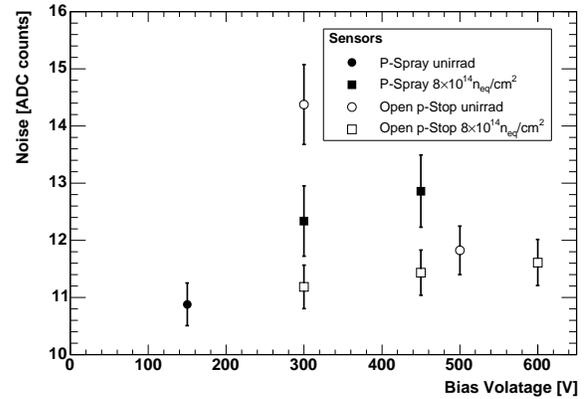

Fig. 5. Bias dependence of the noise on four bump bonded pixel sensors. The central value indicates the mean and the error bar the sigma of a Gaussian fit to the noise distribution of all connected pixels in each sensor.

### B. Characteristics and Noise

The current vs. voltage characteristic is a very sensitive tool used in detecting possible problems in a sensor. Especially after irradiation an early current increase is an indication for electrical breakdown. Early breakdown accompanied by drastic noise increase was previously observed in irradiated *p*-stop isolated devices [13], [19] and is considered to be their major drawback.

To improve the breakdown behavior of the *p*-stop devices, the implantation dose of the *p*-stop implant has systematically been reduced from the typically used $1.2 \times 10^{14}\,\mathrm{cm^{-2}}$ down to the *p*-spray level of $3.6 \times 10^{12}\,\mathrm{cm^{-2}}$. The IV-characteristics of a set of those devices after a proton irradiation of $\Phi = 8 \times 10^{14}\,\mathrm{n_{eq}/cm^2}$ are plotted in Fig. 4 together with a *p*-spray device. It can be seen, that the current and especially the slope of the curve decrease with decreasing *p*-stop dose. A distinct improvement is especially achieved when the *p*-stop dose decreases below $10^{14}\,\mathrm{cm^{-2}}$, while the curves of the devices with an implant dose of $3.6 \times 10^{13}\,\mathrm{cm^{-2}}$ and $1.2 \times 10^{13}\,\mathrm{cm^{-2}}$ hardly differ. The sensors with a *p*-stop dose of $3.6 \times 10^{12}\,\mathrm{cm^{-2}}$ show an IV-characteristic similar to the *p*-spray sensor. If a large number of devices is tested there is a variation in the IV-curves for each implantation dose and it is possible that a sensor with a high *p*-stop implantation dose shows a good IV-curve after irradiation. However the probability of failure because electrical breakdown is strongly reduced if the implantation dose is below about $5 \times 10^{13}\,\mathrm{cm^{-2}}$.

In total, six sensors of the two designs have been bump bonded to readout chips: one of each design unirradiated and one of each design irradiated to $8 \times 10^{14}\,\mathrm{n_{eq}/cm^2}$. The *p*-stop implantation dose of this irradiated sensor was $3.6 \times 10^{13}\,\mathrm{cm^{-2}}$. In addition there was one open *p*-stop sensor irradiated to $3 \times 10^{14}\,\mathrm{n_{eq}/cm^2}$ and a *p*-spray sensor irradiated to $1.1 \times 10^{15}\,\mathrm{n_{eq}/cm^2}$. For the comparison of the designs the unirradiated sensors and the ones irradiated to $8 \times 10^{14}\,\mathrm{n_{eq}/cm^2}$ were used.

Since a full non zero-suppressed readout is possible, the noise of each pixel can be easily extracted from the test beam data. The irradiated devices were operated at $-20^o$ C while the unirradiated were run at $-10^o$ C. In Fig. 5 the bias dependence of the average pixel noise is shown for two sensors of each type unirradiated and irradiated to $8 \times 10^{14}\,\mathrm{n_{eq}/cm^2}$. 12 ADC-counts correspond to a noise of about 400 electrons. This high value is caused by the special chip settings. The noise of the readout chip alone, measured at unbonded pixels, is 10.5 - 11 ADC counts.

For the irradiated sensors no distinct bias dependence of the noise is observed. The noise of the irradiated *p*-spray sensors is about 1 ADC count higher than the noise of the open *p*-stop sensor, which is not significant.

The additional *p*-spray sensor irradiated to $1.1 \times 10^{15}\,\mathrm{n_{eq}/cm^2}$ (not plotted in Fig. 5) shows noise of 17.4 ADC counts at 600 V. The reason for this high noise is not completely understood as only part of it can be explained by the sensor's high leakage current ($36\,\mu$A at 600 V and $-20^o$ C) which occurred after the bump bonding procedure and is very likely due to the mechanical stress. However this sensor worked very stable and reliably. Noise and signal height showed only small variations from pixel to pixel.

The noise of the unirradiated open *p*-stop sensor at 300 V is quite high (14.4 ADC counts) and decreases with increasing bias to 11.8 ADC counts at 500 V. This is because the inter-pixel resistance increases and the coupling between pixels becomes weaker.

The error bars in Fig. 5 represent the sigma of a Gaussian fit to the noise distribution in each sensor and therefore the variation of the noise within the pixels of the sensor. In no case the width of the distribution increases significantly with bias. Further we observe no localized noisy regions which would also lead to an increase in the width of the noise distribution.

From the absence of noisy regions and from the shape of the IV-curves we conclude that electrical breakdown in *p*-stop isolated silicon detectors can be avoided by reducing the implantation dose below roughly $5 \times 10^{13}\,\mathrm{cm^{-2}}$.

### C. Charge Collection Properties

The high energy pion beam of the CERN SPS (150-225 GeV) together with the high precision beam telescope [17]

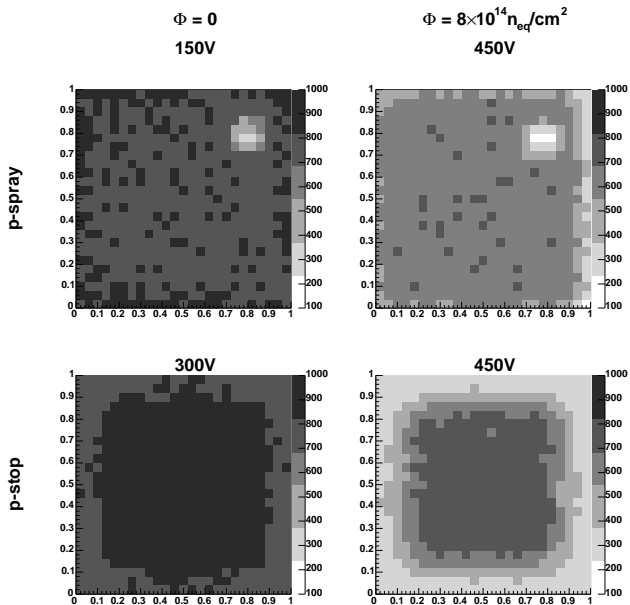

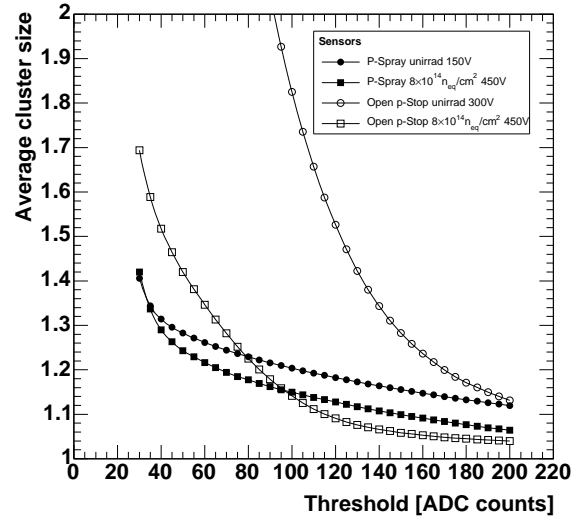

Fig. 6. Cluster charge as function of the pion point of incidence. The area shown represents one pixel cell of $125 \times 125 \,\mu m^2$, the charge is given in ADC counts. The $p$-spray (above) and the open $p$-stop (below) designs are shown for different bias voltages. Irradiation fluence is $\Phi = 0$ (left) and $\Phi = 8 \times 10^{14} \, n_{eq}/cm^2$ (right).

Fig. 7. Average cluster size for the different sensors as function of the pixel threshold.

allows detailed studies of the pixel sensor. Of particular interest was the charge collection behavior as a function of the particle impact point in order to locate "blind" spots within a pixel cell.

Fig. 6 shows the mean total cluster charge deposited by perpendicular tracks as a function of the pion impact position. The area shown represents one square pixel cell with a pitch of $125 \,\mu m$. The cluster signal was obtained by summing the signals of the $3 \times 3$ pixels around the impact point (pixel threshold: 20 ADC counts, cluster threshold: 50 ADC counts). The average amplitude of the cluster signals are listed in table I. In addition, values are given for the signal amplitude in the pixel center and regions with reduced charge collection.

The cluster size was determined by counting the number of pixels above threshold in the direct neighborhood of the impact point. Its average value as a function of the applied pixel threshold is plotted in Fig. 7. The average cluster size for a threshold of 60 ADC counts is listed in table I.

To calculate the signal over noise ratio of a pixel, only the signal in the "hit pixel", the pixel pointed to by the beam telescope, is taken into account. It is obtained by histogramming the charge in this pixel and calculating its mean value, which is listed in table I. Values are also given for tracks passing the pixel center[4]. The signal to noise ratio was calculated by dividing the average signal in the hit pixel by its noise. This procedure, different from the common division of the cluster signal by the noise of one channel, is more suited to describe the performance of the sensor in a zero suppressed system like the final CMS pixel detector. The resulting number gives direct information on the safety margin available for the threshold adjustment.

*1) Unirradiated p-spray sensors:* For unirradiated $p$-spray sensors a very homogeneous average cluster signal of about 800 ADC counts is observed. At the position of the bias dot[5], it drops to less than half of this value. As this area represents only 2–3 % of the total surface, the average collected signal is only weakly affected.

On average about 89 % of the charge is collected by the hit pixel. For tracks in the pixel center this number increases to 96 %. Consequently, the average cluster size is below 1.3 pixels. The large fraction of the cluster signal being collected by one pixel leads to a high signal over noise ratio of 65.

*2) Irradiated p-spray sensors:* After irradiation with a fluence of $\Phi = 8 \times 10^{14} \, n_{eq}/cm^2$, the value of the total collected cluster charge is reduced by about 25 % due to trapping. Furthermore an additional area of reduced charge collection appears at the metal line connecting the bias dots. The collected cluster charge of particles hitting this region[6] is almost 40 % smaller than those hitting the pixel center. As there is no direct contact between this metal line and the silicon below, this behavior (also reported in [20]) was unexpected and is not yet fully understood. It seems plausible that the charge loss is caused by capacitive coupling. Before irradiation the metal line is shielded by the conductive $p$-spray layer. However, after irradiation the "highly" doped part of the $p$-spray layer close to the surface is depleted by the irradiation induced fixed surface charge and behaves like a dielectric. The undepleted tails of the boron profile have nearly intrinsic conductivity and behave like a (bad) insulator. Signal charge drifting below the $p$-spray layer can induce an electrical signal on the bias line above. Since the total affected area is under

---

[4]The region of $0.26 < x < 0.74$ and $0.36 < y < 0.64$. It does not show significant variations in the collected signal in any sensor.

[5]The region between $0.76$ and $0.84$ in $x$ and $y$. This is the position of the bias dot visible in Fig. 1b.

[6]Region with $y > 0.96$.

TABLE I
SUMMARY OF TEST BEAM RESULTS.

| $\Phi$ [$n_{eq}/cm^2$] | Bias [V] | Mean Cluster Signal [ADC] Average | Center | Border | Signal in hit pixel [ADC] Average | Center | Cluster Size at 60 ADC | Noise [ADC] | S/N | Efficiency at 60 ADC |
|---|---|---|---|---|---|---|---|---|---|---|
| | | | | *p*-spray | | | | | | |
| 0 | 150 | 800 | 803 | dot: 366 | 710 | 768 | 1.26 | 10.9 | 65 | 99.69 % |
| $8 \times 10^{14}$ | 450 | 599 | 647 | dot: < 250 line: 342 | 533 | 612 | 1.22 | 12.9 | 41 | 98.56 % |
| | | | | open *p*-stop | | | | | | |
| 0 | 300 | 847 | 905 | 765 | 424 | 569 | 3.36 | 14.4 | 29 | 99.67 % |
| $8 \times 10^{14}$ | 450 | 539 | 718 | 323 | 465 | 700 | 1.35 | 11.4 | 41 | 99.32 % |

10 %, the signal averaged over the whole pixel cell is about 7 % smaller than the signal collected in the central pixel region.

The charge loss at the position of the bias dot is more severe than in the unirradiated sensor. As the applied thresholds already cut into the lower part of the cluster spectrum, we quote the average as an upper limit.

The charge sharing between pixels is not effected by the irradiation. Still 89 % of the cluster charge is collected by one pixel and the average cluster size decreases only a little. As the signal height decreases and the noise slightly increases, the signal over noise ratio of this irradiated sensor decreases to 41.

For the additional *p*-spray sensor irradiated to $\Phi = 1.1 \times 10^{15}\,n_{eq}/cm^2$ and operated at 600 V, the general behavior remains unchanged. However, due to the cluster signal further reduced by trapping to an average value of 533 ADC counts and the increased noise, the signal over noise ratio is further reduced to 26. A discussion of the charge collection efficiency as a function of the particle fluence and the bias voltage and the drift length of the signal charge is presented in [18].

*3) Unirradiated p-stop sensors:* The highest average cluster signals were observed in the center of an unirradiated open *p*-stop sensor, 905 ADC counts, which is about 12 % larger than the maximum in the *p*-spray devices. However, one has to take into account the use of 300 V bias voltage in order to reduce the charge spread due to the resistive connections between pixels. Even at such a high bias voltage the average cluster size, using a pixel threshold of 60 ADC counts, is above 3, much higher than in the other investigated sensors. Even if a pixel is hit in the center, it carries only 63 % of the total cluster signal (in average it is only 50 %). Close to the pixel border[7], the cluster signal decreases by about 15 % compared to the central region. As the hit pixel carries only a small fraction of the total cluster signal, the signal to noise ratio is only 29.

*4) Irradiated p-stop sensors:* Due to the irradiation induced increase of the inter-pixel resistance, the average cluster size of the irradiated open *p*-stop sensors decreases to values below 1.5. At the same time the fraction of the cluster signal collected by the hit pixel increases to 85 %, the same level as in the *p*-spray sensors. While the average cluster charge in the sensor irradiated to $\Phi = 8 \times 10^{14}\,n_{eq}/cm^2$ decreases by about 20 % compared to the unirradiated sensor, the charge collected by the hit pixel stays about unchanged. When the pixel is hit in its center, the charge collected by the hit pixel even increases.

[7]Region with $x$ or $y$ < 0.08 or > 0.92.

The concentration of signal charge on one pixel leads to an increase of the signal over noise ratio to 41, the same value as the *p*-spray sensor irradiated to the same fluence.

The cluster charge for tracks close to the pixel border decreases to half of the value for tracks in the center. The reason for this significant charge loss is not fully understood but the following explanation seems possible: The electron accumulation layer between the open *p*-stops adjusts to the same potential as the pixel implants due to the openings in the open *p*-stops. For this reason the layer also collects signal charge. For the unirradiated sensor the surface mobility in the electron accumulation layer is high enough to allow a quick transfer of the collected charge to the next readout $n^+$-implant. Therefore the average cluster size in the unirradiated sensor is very high. After irradiation the mobility of free charge carriers close to the surface is strongly reduced and the number of surface traps increases. The charge drift to the next readout $n^+$-implant is slower and a significant fraction of the signal collected in the accumulation layer might not reach it in time. Hence the cluster size in the irradiated open *p*-stop sensor is much smaller.

### D. Particle Detection Efficiency

In the operation of the pixel detector, an important figure of merit is the probability for detecting a particle penetrating the detector. As the final readout chip will feature a single threshold zero suppression and only pixels exceeding this threshold can be read out, a particle can be counted as detected, if it triggers at least one pixel. In order to translate the charge collection behavior discussed in the previous sections into efficiencies, realistic thresholds have been applied. If the pixel pointed to by the beam telescope or a direct neighbor was above the threshold, the track was counted as detected. Regions of defect bump bonding or noisy pixels were excluded from the analysis.

The noise of all sensors is between 11 and 13 ADC counts, a threshold of 60 ADC counts, about five times the noise, was chosen. This value corresponds to a signal charge between 2000 and 2200 electrons and is close to the expected value during LHC operation. The probability for a particle triggering a readout with a pixel threshold of 60 ADC channels is given in table I. In all cases it is above 98 %.

*1) P-spray sensors:* The inefficiency as a function of the threshold is plotted in Fig. 8 for both designs under study, and for proton fluences of 0 and $8 \times 10^{14}\,n_{eq}/cm^2$. In the unirradiated *p*-spray design the charge loss due to the bias dot is small enough not to cause an inefficiency if the pixel threshold is below 100 ADC counts. For higher thresholds the lost tracks start to concentrate around this area. In the case of the irradiated $p$-spray sensor the probability for loosing a track hitting the bias dot is higher due to the lower total charge. Already at low thresholds the total inefficiency is about 1 %. With increasing threshold the inefficiency rises as well. Tracks hitting the region below the metal line of the bias grid start to contribute, beginning from the corner of the pixel and growing along the pixel edge below the metalization. When the threshold exceeds 130 ADC channels, a small accumulation of lost tracks can also be found in the corners opposite to the metal line. As the total affected area is quite small, the slope of the increase is limited and the inefficiency does not exceed 4 % even at high thresholds (e.g. 160 ADC counts).

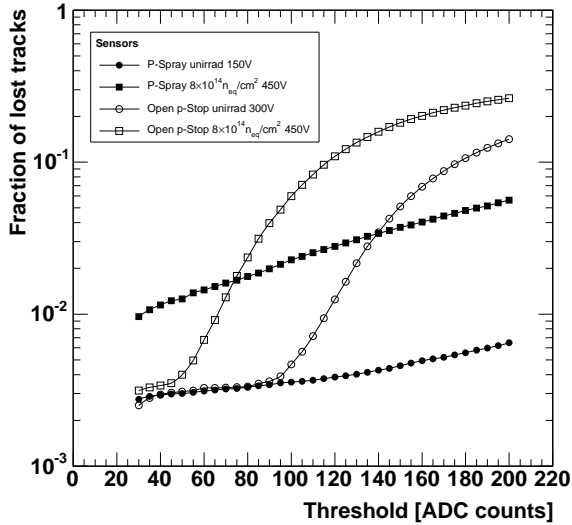

Fig. 8. Fraction of undetected tracks in the pixel detector as function of the applied threshold.

*2) Open p-stop sensors:* For the open $p$-stop designs the situation looks different. The charge loss in the pixel edge region is less drastic and the efficiency at low thresholds is above 99.5 %, which is also true for the irradiated sensor. However, as the less efficient region at the pixel edge covers a significant fraction of the area, the inefficiency increases rapidly with increasing threshold. The lost tracks accumulate at the pixel corners. With increasing threshold the regions of lower efficiency grow along the pixel edges.

Although the irradiated open $p$-stop sensor reach a better efficiency at a threshold of 60 ADC counts, it has to be stressed that the high slope of the inefficiency displays a potential risk. A small threshold variation can lead to a non-tolerable inefficiency above 5 %. The open $p$-stop sensor irradiated to $8 \times 10^{14}\,\mathrm{n_{eq}/cm^2}$ has also been measured at a bias voltage of 600 V. The higher bias increases the collected charge and the detection efficiency, although not significantly.

The situation in the test beam with perpendicular tracks without magnetic field is well suited for characterizing the charge collection properties of the sensors with high precision. In the final experiment, however, the signal charge will always be spread over a certain area due to the track inclination and the Lorentz drift. Therefore the efficiency numbers given in this section cannot be directly applied. The effect of small regions of reduced charge collection like the bias dot in the $p$-spray sensor and the metal line, which will be oriented perpendicular to the Lorentz drift, will probably be suppressed. For the open $p$-stop sensor, where this regions occupy a significant fraction of the pixel area, the steep increase of the inefficiency might effect the detector performance noticeably.

## V. CONCLUSIONS

Silicon pixel sensors of "$n$-in-$n$" type featuring $p$-spray and open $p$-stop isolation have been irradiated up to proton fluences of $1.1 \times 10^{15}\,\mathrm{n_{eq}/cm^2}$. All sensors show IV-curves with a breakdown voltage well above 600 V without localized noisy regions. In the case of open $p$-stop sensors this was achieved by reducing the open $p$-stop implantation dose to about $10^{13}\,\mathrm{cm^{-2}}$.

The charge collection studies were performed with bump bonded samples using a high energy pion beam. The total charge collected after the highest fluence applied was about 60 % of the value obtained with unirradiated sensors independent of the sensor design. The main results for the different sensor type are the following.

1) P-spray sensors:
   - The $p$-spray devices showed a very homogeneous charge collection also in the inter-pixel regions.
   - The bias dots represent an area with strongly reduced charge collection, leading to a loss of particle detection efficiency of about 1 % after an irradiation fluence of $8 \times 10^{14}\,\mathrm{n_{eq}/cm^2}$.
   - For irradiated sensors the metal line of the bias grid additionally reduces the charge collection.
   - The particle detection efficiency after this fluence still exceeds 98 % and is only moderately dependent on the pixel threshold.
2) Open p-stop sensors:
   - In the unirradiated sensor the signal is spread over many pixels.
   - After irradiation this spread is strongly reduced.
   - The most inefficient region is located in between the pixels. The cluster signal of a track hitting an irradiated sensor close to the pixel border is only half of the size of a central hit.
   - The particle detection efficiency is above 99 % at low threshold but drops drastically for thresholds higher than about 95 ADC counts for the unirradiated sensor and 55 ADC counts for the sensor irradiated to $8 \times 10^{14}\,\mathrm{n_{eq}/cm^2}$.

The steep increase of lost tracks seems to be the major drawback of the open $p$-stop sensors and has to be further investigated. A possible improvement of the charge collection in the inter-pixel region might be possible, if the pattern of the resistive inter-pixel connections is changed. Every pixel should not be coupled resistively to an overall network as indicated in Fig. 2, but only to its direct neighbors. In this case also the large spread of the signal charge in the unirradiated sensor will be reduced. Such devices were already built and will be investigated in the near future.

The *p*-spray sensors could be used in CMS without further modification.

ACKNOWLEDGMENT

The authors would like to thank Silvan Streuli from ETH Zürich and Fredy Glaus from PSI for their enormous effort in bump bonding, Kurt Bösinger from the workshop of the University of Zürich for the mechanical construction, Maurice Glaser and Michael Moll from CERN for carrying out the irradiation, and György Bencze and Pascal Petiot from CERN for the H2-beam line support. Thanks are also due to Danek Kotlinski for the very useful discussions and for sharing his detailed knowledge of the pixel control and readout system. Last but not least we gratefully acknowledge Roland Horisberger from PSI for explaining all details of the PSI30/AC30 readout chip. Without his advice this work would not have been possible.